\documentstyle[12pt]{article}
\catcode`\@=11
\def\maketitle{\par
\begingroup
\def\thefootnote{\fnsymbol{footnote}}
\def\@makefnmark{\hbox
 to 0pt{$^{\@thefnmark}$\hss}}
\if@twocolumn
\twocolumn[\@maketitle]
\else \newpage
\@maketitle \fi
\thispagestyle{empty}
\@thanks
\endgroup
\setcounter{footnote}{0}
\let\maketitle\relax
\let\@maketitle\relax
\gdef\@thanks{}
\gdef\@author{}
\gdef\@title{}
\gdef\@preprint{}
\let\thanks\relax}

\def\@maketitle{
\hbox to\textwidth{\hfil
\hbox{\begin{tabular}{l}\@preprint\end{tabular}}}
\vskip 2em
\begin{center}{\Large\bf \@title \par}
\vskip 1.5cm {\normalsize
\lineskip .5em
\begin{tabular}[t]{c}\@author
\end{tabular}\par}
\end{center}\par\vskip 1.5em}

\def\preprint#1{\gdef\@preprint{#1}}

\def\abstract{\if@twocolumn
\section*{Abstract}
\else \normalsize
\begin{center}
{\large\bf Abstract\vspace{-.5em}\vspace{0pt}}
\end{center}\quotation\fi}
\def\endabstract{\if@twocolumn\else\endquotation\fi}
\catcode`\@=12

\begin{document}
\baselineskip 6mm
\preprint{DPNU-96-52 \\
September 1996}
\title{{\bf Negative dimensional group extrapolation \\ 
and \\
dualities in $N=1$ supersymmetric gauge theories}}
\author{Nobuhito Maru\thanks{e-mail address: 
maru@eken.phys.nagoya-u.ac.jp} and Shinsaku 
Kitakado\thanks{e-mail address: g44332a@nucc.cc.nagoya-u.ac.jp} \\
\\
{\em Department of Physics, Nagoya University} \\
{\em Nagoya 464-01, Japan}}
\date{}
\maketitle
\setcounter{page}{0}
\thispagestyle{empty}
\begin{abstract}
We point out that the similarities in $N = 1$ 
supersymmetric $SO$, $SP$ gauge theories 
can be explained by
using the trick of extrapolating the groups 
to the negative dimensions.
One of the advantages of this trick is 
that anomaly matching is automatically satisfied.
\end{abstract}
\newpage

\baselineskip 18pt
\section{Introduction} 
\indent 


In a couple of years, the understanding of 
non-perturbative properties 
of supersymmetric gauge theories 
has been rapidly enlarged. 
Especially, duality proposed by Seiberg \cite{Sei95} 
gives a clue 
to analyze the strong interaction dynamics. 
Many authors \cite{ILS} 
have suggested dualities which have various gauge groups 
( $SU,SO,SP$ group and their product group ), 
various matter contents
(at most 2-index tensor ).

In this paper we point out that the 
similarities of dualities 
in $N = 1$ supersymmetric $SO$, $SP$ 
gauge theories can be 
explained by using the trick of negative 
dimensional groups. 
In constructing the duality in $N = 1$ 
supersymmetric gauge 
theories one of the key idea is anomaly matching. 
Anomaly matching conditions heavily 
depend on the representations 
and the charges of matter fields in the model. 
Therefore we have 
to find the matter content of the model by trial and error. 
However if we know one $SO(SP)$ duality 
we can easily obtain 
another $SP(SO)$ duality which automatically 
satisfies anomaly 
matching conditions by using the trick of negative 
dimensional groups.

The idea of negative dimensional groups is not new and 
goes back to Penrose \cite{Penrose} 
who has constructed the $SU(2)$ 
( = $SP(2)$ ) representations in terms of $SO( - 2 )$. 
Since then many relations have been 
observed among the expressions for 
the $SU(N)$, $SO(N)$ and $SP(N)$ 
group invariants under the 
substitution $N \to - N$ \cite{Negadim}. 
On the other hand, Parisi and Sourlas \cite{Parisi} 
have observed 
that a Grassmann space of dimension $N$ 
can be interpreted as an ordinary 
space of dimension $ - N $. 
In supersymmetric theories the $N \to - N$ relations are, 
in a sense, built in and we are going to utilize 
this property in this paper.

\section{Similarities in $SO$ and $SP$ duality} 
\indent


In order to show the advantages of our 
trick of negative dimensional groups 
we focus here on dualities with $SO$ and $SP$ 
gauge groups
\footnote{As will be explained later, $SU(N)$ 
group is self-dual under 
$N \to - N$. Therefore we don't 
discuss $SU$ group here.}. 
It is well known that these models strongly 
resemble each other in 
appearance. 
As an example, we shall take the model 
proposed by Intriligator \cite{KI}. 
In Ref.\cite{KI}, duality in supersymmetric $SO$ and $SP$ 
gauge theories 
are discussed. The electric theory of $SO$ 
dual model is a $N=1$ 
supersymmetric $SO(2 N_c)$\footnote{In order to see 
the relation with 
the $SP$ groups we are restricting our 
discussion to the even dimensional $SO(2 N_c)$ 
groups leaving aside the $SO(2 N_c + 1)$ groups} 
gauge theory with $2 N_f$ fields $Q^i$ 
in the fundamental representation and 
a symmetric traceless 
tensor $X$. The anomaly free global symmetries are 
$SU(2 N_f) \times U(1)_R$ with 
the fields transforming as

\begin{eqnarray}
Q^i && \left( 2 N_f, 1 - \frac{2 (N_c - k)}{(k + 1) N_f} \right), 
\nonumber \\
X && \left(1, \frac{2}{k + 1} \right).
\end{eqnarray}
The superpotential is 

\begin{equation}
W = g_k Tr X^{k + 1}.
\end{equation}

The magnetic theory is $N=1$ supersymmetric $SO(2 \tilde{N_c})$ 
gauge theory, where $\tilde{N_c} \equiv k (N_f + 2) - N_c$, 
with $2 N_f$ fields $q^i$ in the fundamental representation, 
a symmetric traceless tensor $Y$ and 
singlets $M_j (j = 1, \cdots, k)$. 
The anomaly free global symmetries 
are $SU(2 N_f) \times U(1)_R$ with 
the fields transforming as 

\begin{eqnarray}
q^i && \left( \overline{2 N_f}, 1 - 
\frac{2 (\tilde{N_c}-k)}{(k+1)N_f} 
\right), \nonumber \\
Y && \left( 1, \frac{2}{k+1} \right), \\
M_j && \left( N_f(2 N_f + 1), \frac{2(j+k)}{(k+1)} - 
\frac{4(N_c-k)}{(k+1)N_f} \right). \nonumber
\end{eqnarray}
The superpotential is 

\begin{equation}
W = Tr Y^{k+1} + \sum^k_{j=1} M_j q Y^{k-j} q.
\end{equation}

On the other hand, the electric theory of $SP$ dual model 
is a $N=1$ 
supersymmetric $SP(2 N_c)$ gauge theory 
with $2 N_f$ fields $Q^i$ in the fundamental 
representation and an antisymmetric 
traceless tensor $X$
\footnote{In this paper, we denote the symplectic 
group as $SP(2 N_c)$
whose fundamental representation is $2N_c$ dimensional.}. 
The global symmetries are $SU(2 N_f) \times U(1)_R$ 
with fields transfoming as

\begin{eqnarray}
Q^i && \left( 2 N_f, 1 - \frac{2 (N_c + k)}{(k + 1) N_f} \right), 
\nonumber \\
X && \left(1, \frac{2}{k + 1} \right).
\end{eqnarray}
The superpotential is 

\begin{equation}
W = g_k Tr X^{k + 1}.
\end{equation}

The magnetic theory is $N=1$ supersymmetric $SP(2 \tilde{N_c})$ 
gauge theory, where $\tilde{N_c} \equiv k (N_f - 2) - N_c$, 
with $2 N_f$ fields $q^i$ in the fundamental representation, 
an antisymmetric traceless tensor $Y$ 
and singlets $M_j (j = 1, \cdots, k)$. 
The global symmetries are 
$SU(2 N_f) \times U(1)_R$ with the fields transforming as 

\begin{eqnarray}
q^i && \left( \overline{2 N_f}, 1 - \frac{2 (\tilde{N_c}+k)}{(k+1)N_f} 
\right), \nonumber \\
Y && \left( 1, \frac{2}{k+1} \right), \\
M_j && \left( N_f(2 N_f - 1), \frac{2(j+k)}{(k+1)} - 
\frac{4(N_c+k)}{(k+1)N_f} \right). \nonumber
\end{eqnarray}
The superpotential is 

\begin{equation}
W = Tr Y^{k+1} + \sum^k_{j=1} M_j q Y^{k-j} q.
\end{equation}
It is easy to recognize that the representations 
and the charges of fields 
are quite similar. Futhermore, it can be seen 
that we can obtain 
$SP(SO)$ duality from the $SO(SP)$ duality by changing 
the signs of $N_c$, $N_f$ into $-N_c$, $-N_f$ and 
exchanging a symmetric 
(an antisymmetric) tensor for an 
antisymmetric (a symmetric) tensor. 
This feature is not specific to these models 
and is applicable to 
$SO, SP$ dual models discovered so far \cite{ILS}.

\section{Negative dimensional group}
\indent

Group theoretically, these can be anticipated 
by considering the 
negative dimensional groups first proposed 
by Penrose \cite{Penrose}. 
This is a technique to calculate 
the algebraic invariants.
Using this technique, we can find the 
peculiar relations for 
dimensions of the irreducible representations of 
the classical groups 
$SU(N), SO(N), SP(N)$ \cite{Negadim}. 
If $\lambda_s$ is a Young tableau 
with s boxes and if the dimensions of 
the correponding irreducible 
representations of $SU(N), SO(N)$ and $SP(N)$ 
are denoted by 
$D \{\lambda_s ; N \}$, $D[\lambda_s ; N]$ and 
$D\langle \lambda_s ; N \rangle$, respectively, 
it was noticed by King \cite{Negadim} that 

\begin{eqnarray}
D \{\lambda_s ; N \} &=& (-1)^s 
D \{\tilde{\lambda_s} ; - N \}, \\
D[\lambda_s ; N] &=& (-1)^s 
D \langle \tilde{\lambda_s} ; - N \rangle.
\end{eqnarray}
Here $\tilde{\lambda}$ stands for 
the "transposed" (rows and columns 
interchanged ) Young tableau.
Moreover, it is useful to give the relations 
among the generalized 
Casimirs of the classical groups in totally symmetric and 
totally antisymmetric representations.

\begin{eqnarray}
C^{SU(N)}_p (1,1,\cdots,1) &=& (-1)^{p-1} C^{SU(-N)}_p 
(r,0,\cdots,0), \\
C^{SO(2N)}_p (1,1,\cdots,1) &=& (-1)^{p-1} C^{SP(-2N)}_p 
(r,0,\cdots,0), \\
C^{SP(2N)}_p (1,1,\cdots,1) &=& (-1)^{p-1} C^{SO(-2N)}_p 
(r,0,\cdots,0),
\end{eqnarray}
where $C_p (1,1,\cdots,1)$ and $C_p (r,0,\cdots,0)$ 
mean the p-th 
order generalized Casimir in totally antisymmetric and 
totally symmetric rank-r tensor representations respectively. 
These relations Eqs.(9)-(13) are necessary to 
take anomaly matching conditions into account.

We can express these results symbolically 
as follows \cite{Negadim},

\begin{eqnarray}
SU(-N) &\cong& \overline{SU(N)}, \nonumber \\
SO(-N) &\cong& \overline{SP(N)}, \\
SP(-N) &\cong& \overline{SO(N)}, \nonumber
\end{eqnarray}
where the overbar means symmetrization and antisymmetrization 
are interchanged.

Since the supersymmetric theories are "invariant" 
under this interchange 
we can use this technique as a useful method of 
obtaining a dual model from 
another dual model through the extrapolation 
to the negative 
dimensional groups. The procedures are :
\begin{itemize}
\item Change the sign of the group dimension, 
$N_c \leftrightarrow -N_c$, $N_f \leftrightarrow -N_f$. \nonumber \\
\item Interchange the symmetrization 
and antisymmetrization of the 
representations. \nonumber
\end{itemize}
We can actually convince ourselves 
that with these procedures 
Eq.(7) follows from Eq.(3). 
This negative dimensional group technique is very powerful since 
the anomaly matching is automatically satisfied.

\section{Summary and Discussion} 
\indent 


In this paper we have pointed out that the similarities 
of dualities in $N = 1$ supersymmetric $SO$, $SP$ 
gauge theories 
can be explained by using the trick of 
negative dimensional groups. 
If we know the duality in supersymmetric $SO(SP)$ gauge 
theory we can easily obtain another duality in 
supersymmetric $SP(SO)$ gauge theory which automatically 
satisfies the anomaly matching conditions by extrapolating 
the groups to the negative dimensions in the model. 
By this trick we can also know the representations and 
the charges of the fields. On the other hand, 
when there is no known duality this trick is powerless.

Explicit application of this trick in finding new dualities 
is left for future studies. 
It is interesting to see whether this 
trick is applicable to other groups. 
We have described the substitution $N \to N$ just as 
a useful trick for studying the duality structures of 
supersymmetric theories leaving aside the direct significance 
it might have in such theories. Thus it can be interesting 
to study the symmetries under $N \to N$ directly in the 
supersymmetric theories where duality is realized explicitly. 
We hope to report elsewhere these 
together with the related problems. 

\vspace{1cm}

\noindent{\bf Acknowledgement}

\vspace{0.2cm}

We thank E.Poppitz, P.Pouliot and M.J.Strassler for valuable 
comments and for pointing out our mistakes.


\end{document}